# A Feature Based Methodology for Variable Requirements Reverse Engineering

**Anas Alhamwieh, Said Ghoul***

Faculty of Information Technology, Research Laboratory on Bio-inspired Software Engineering, Philadelphia University, Amman, Jordan

**Email address:**
anas.alhamyea@gmail.com (A. Alhamwieh), sghoul@philadelphia.edu.jo (S. Ghoul)
*Corresponding author

**Abstract:** In the past years, software reverse engineering dealt with source code understanding. Nowadays, it is levered to software requirements abstract level, supported by feature model notations, language independent, and simpler than the source code reading. The recent relevant approaches face the following insufficiencies: lack of a complete integrated methodology, adapted feature model, feature patterns recognition, and Graph based slicing. This work aims to provide some solutions to the above challenges through an integrated methodology. The following results are unique. Elementary and configuration features are specified in a uniform way by introducing semantics specific attributes. The reverse engineering supports feature pattern recognition and requirements feature model graph-based slicing. The slicing criteria are rich enough to allow answering questions of software requirements maintainers. A comparison of this proposed methodology, based on effective criteria, with the similar works, seems to be valuable and competitive: the enrichment of the feature model and feature pattern recognition were never approached and the proposed slicing technique is more general, effective, and practical.

## 1. Introduction

One of the main Software Product Lines (SPLs) purposes is to identify and to manage the variations between the whole products family requirements; these variants provide different functional and non functional requirements based on features [1]. SPLs are defined as a family of systems that share a common set of core technical assets, with a pre-planned extensions and variations to address the requirements of specific customers and market [2]. Several methods have been used to manage and formalize the variability among products requirements. Feature Models (FM) are the most popular ones. Their major aspect is a graph representation that includes a group of defined features and relations between them [3, 4, 5]. Management operations have been established for reverse engineering, merging, slicing, or refactoring feature models from a group of configurations [3].

One important process, that should be mentioned, in software engineering is reverse engineering [19, 20]. Certain authors have defined it as the process of analyzing a subject system to identify the system's constituents and creating representations in another form at higher levels of abstraction [6]. These authors show how meta-models are frequently used, during source code based reverse engineering. Another research work dealt with the challenge of understanding and analyzing parts of software using a reverse engineering tool that clarifies the limitations of all past source code based reverse engineering [7]. Software reverse engineering is mainly supported by patterns recognition, slicing, and shopping techniques. Different works presented pattern recognition in different domains [9, 10]. However, there is no works on feature pattern nor on feature pattern recognition in FM. Several works presented feature model slicing techniques and algorithms based on mathematical formula model [13, 14]. But, no one is based on feature graph model itself; which is a conceptual weakness.

Software requirements are becoming a specialized engineering science [21, 22]. It is supported by engineering methodologies, methods, techniques, and tools. But, so far, it suffers of the lack of reverse engineering fundamentals. This is mainly due to the recent accreditation of the requirements engineering and to the non-formal nature of its large products number. Anyway, the requirements reverse engineering should help in requirements understanding by successive abstractions and meta models. This understandability is the kernel of any requirements evolution. The idea of feature-based requirements reverse engineering is similar to program reverse engineering [6, 7]. They differ only on used techniques. Feature model slicing is inspired from program slicing that is currently used in computer programming to avoid and eliminate all parts of the program that are not relevant to a current interest [11, 12].

All of the research works on FM reverse engineering do not deal with (1) FM notations suitability to support reverse engineering tasks, (2) features patterns recognition, (3) graph-based slicing, and a methodology uniformly integrating the above activities [15, 16, 17]. This work overcomes the above stated shortages. It unifies the definition of an elementary feature as well as the composed (configuration) one by enhancing the feature notation with its semantics. It introduces feature pattern recognition allowing the understanding of any requirement feature semantics, and the feature model graph-based slicing allowing the identification of (1) a sub feature model which might affect a given requirement feature (backward slice) or (2) a sub feature model which might be affected by a given requirement feature (forward slice). The proposed methodology integrates the above techniques.

A comparison of this work, based on effective criteria, with the similar works, seems to be valuable and competitive: the enrichment of the feature model and feature pattern recognition were never approached and the proposed slicing technique is more general, effective, and useful to requirements maintainer because it is based on graph notations.

## 2. A Methodology for Variable Requirements Reverse Engineering

### 2.1. Supporting Example

The following example, which is a variable requirements FM of a software List (figure 1), will be used along with the proposed methodology. This feature model (ExFM) is an extension of the conventional one. It is composed of two sub models: Elementary Feature Model (EIFM), modeling elementary requirements features, and Evolution Feature Model (EVFM), modeling configurations requirements features (composed by selected elementary or configuration features). The EIFM of List software requirements consists of behaviors requirements, structures requirements, methods requirements, etc. Each feature exists in two variations: static or dynamic. The Requirements Evolution Feature Model (EVFM) of software List requirements deals with List configurations (evolution) aspects.

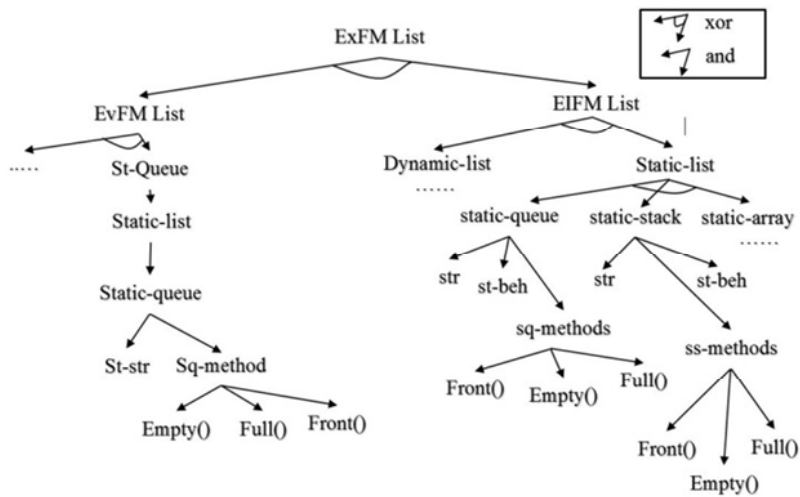

*Figure 1.* Extended FM Diagram of List software [8].

### 2.2. Feature Modelling

This section introduces a uniform definition of elementary and configuration (composed) feature [8]. In the proposed methodology, these two FMs are unified in a unique simplified one. Below the definition of a uniform feature model using EBNF notation [18]:

<FM> = "feature model" < FM name>";"
    < Feature> // no difference between ELFM and EVFM
"End FM" < FM name> ";"

Each feature, in the FM, is composed of attributes and relations:
<Feature> = "feature" <feature name>";"

<Attributes> ";" [<Relations>]

Where:

*Attributes:* define characteristics of a feature:

<Attributes> = "attributes"

((<Attr_name>):<Attr_value>(","<Attr_value>)*)*

- *Relations*: define relations among features:

<Relations> = "relation"

(<decomposition> | <constraint> | <included in>)

The decomposition relations are defined by:

<Decomposition> = "decomposition"

<and> | <xor> | <or> | (select (<feature>)* | default (<feature>)*)*

These relations have the following semantics:

*F2 and F3*: a feature F1 is composed by features F2 and F3 if the two features are compulsory.

*F2 xor F3:* a feature F1 might be composed exclusively by feature F2, or by feature F3 but not by both.

*F2 or F3:* a feature F1 might be composed by feature F2, or by feature F3 or by both.

*Select*: is a relation that determines which features will be selected.

*Default*: is a relation that determines a standard feature to be selected in case where no feature has been explicitly selected.

The constraint relations are defined by:

<Constraint> = "constraints"

<imply> | <exclude> | <reject>

These constraints have the following semantics:

A feature F1 *implies* a feature F2 if a software holding F1 must hold F2 too.

A feature F1 *excludes* a feature F2 if a software holding F1 must not hold F2.

A feature F1 is *rejected* if F1 is unwanted.

The Included in relations have the following semantics:

<Included in> = "included in"

(<Features> ",")*

Figure 2 represents a unified FM of software List (Figure 1) using the semantics introduced above.

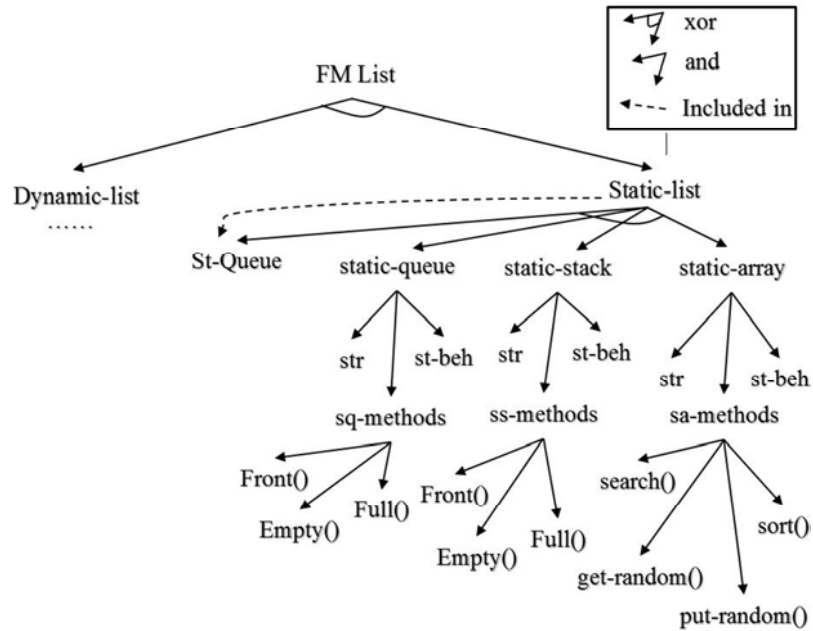

*Figure 2. FM of software list. St-Queue is a configuration feature. The others are elementary ones.*

### 2.3. Reverse Engineering

A FM is composed of elementary and configuration features. Each one of these features is an instance of a predefined pattern specifying its semantics and allowing its.

a   Elementary feature pattern

A pattern of an elementary feature is introduced as it follows:

*FPattern Elementary ( )*

```
    {
      Name: <Feature id>
      Relations:
              Decomposition: [<and> | <xor> | <or>]
              Constraint: [<imply> | <exclude>]
              Included in: [(<Features> ",")*]
    }// end Elementary
```
b   Configuration feature pattern
A pattern for a configuration is proposed to recognize configuration feature:
```
    FPattern Configuration ( )
{
      Name: <Feature id>
      Relations:
      Decomposition: [Select <Feature> (Variation= <Feature>)]* |
                     [Default <feature>] | [Imply <feature>]|
                     [<and> | <xor> | <or>]
      Constraint: [Reject (<feature>)*] | [<imply> | <exclude>]
      Included in: [(<Features> ",")*]
}// end Configuration
```
c   A Feature type mining algorithm
```
    Procedure FeatureTypeMining (in Feature F, out Type T, out Meaning M)
    { // Input: Selected feature F.
    // Output: The type of the selected feature T and its meaning M.
    String FT;
        FT ← Recognize (F); // Call a predefined function Recognize (in feature F) which returns the type of feature F
    ("Elementary" or "Configuration").
        If (FT = "Configuration feature") then
        { T ← Configuration feature;
        M ← (F.Name, F.Decomposition, F.Constraint, F.Included_in)
        }
    Else { T ← Elementary feature;
        M ← (F.Name, F.Decomposition, F.Constraint, F.Included_in)
            }
    }//End FeatureTypeMining
```

## 2.4. Slicing

The slicing of software FM is introduced as it follows:

<p align="center">Slice <Feature> <Direction> <Relation></p>

Where:
   <Feature> is the selected feature in feature model (Elementary or Configuration).
   <Direction> = <Backward | Forward>
   The *Backward direction* identifies the features which might affect the selected feature.
   The *Forward direction* identifies the features (with their different variations) which might be affected by a selected feature.
   <Relations> = (<AND> | <OR> | <Included in>)
   The *AND* relation is defined by < AND > = "AND"
   The *OR* relation is defined by < OR > = "OR" [<(Alternative feature)*>]
   The slice is applied in the same way for both elementary and configuration feature.
The following algorithm find out all the different slices introduced above.
*Procedure FeatureSlicing (in Feature F, in Direction D, in Relation R,*
*[in Alternativefeature A], out FeatureModel FM , out Type T, out Meaning M)*
*{*
   *Input: Selected feature F, Direction D, Relation R, [Alternative feature A].*
   *Output: Feature Model FM, Type of Feature T, Meaning M.*
   *//Call FeatureTypeMining (in Feature F, out Type T, out Meaning M) to return the type T of the feature F and its meaning*

M.
 FeatureTypeMining (F, T, M);
 If (R = "AND") then //AND Relation
 If (D = "Forward") then
 FM ← SelectAND (F) //Call SelectAND (in Feature F) that returns the Feature model slices of F by FM graph traversing (BFS). It is a predefined function.
 Else //Backward
 FM ← Parent (F) // Call Parent (in Feature F) that returns all ancestors for F, it is built in function.
 Else If (D = "Forward") then // OR Relation
 FM ← SelectOR (F, A)) //Call SelectOR (in Feature F, [in Alternativefeature A]) that returns the Feature model slices of F It is built in function.
 Else //Backward
 FM ← Parent (F)
} //End FeatureSlicing

## 2.5. Proposed Methodology

This section introduces the integration of the above proposed techniques through a supporting methodology for variable requirements reverse engineering (Figure 3).

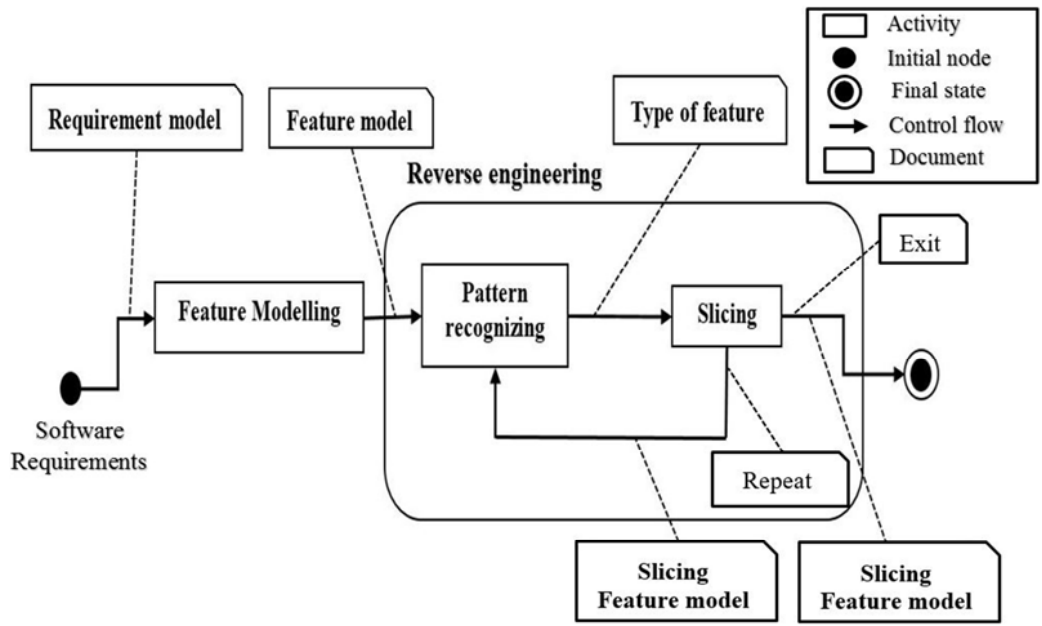

*Figure 3. A Methodology for variable requirements reverse engineering, using UML notation.*

The *Requirement model* is a software functional and non-functional requirements definition using a requirement modelling language. The *Feature Model* is generated from the requirements model. However, it is not adapted to reverse engineering process. Thus, this work enriched it with needed concepts. The *Pattern Recognition* is designed to deal with both elementary feature and configuration feature. The proposed algorithm find out the semantics of a given feature. The *Slicing* process is a graph-based algorithm dealing with side effects of evolution tracking (feature or relation adding, changing, or deleting).

The pattern recognition of the feature St-Queue, in the Feature Model Figure 2, is obtained by the call of the procedure *FeatureTypeMining (in Feature F, out Type T, out Meaning M)*;

Input: F = St-Queue
Output: T = Configuration feature
M = {Name: St-Queue
 Decomposition: Select List (Variation = Static-list, Variation = static-queue);
 Constraint: Reject st-beh
 Included in: ---
 }

The pattern recognition of the feature static_queue, in the Feature Model Figure 2, is obtained by the call of the procedure FeatureTypeMining (in Feature F, out Type T, out Meaning M);

Input: F = static_queue
Output: T = Elementary feature
 M = {Name: static_queue

*Variation: str, st-beh, sq-methods*
*Decomposition: static_queue and (str and st-beh and st-methods)*
*Constraint: static_queue exclude static-stack*
*Included in: St-Queue}*

As side effect to static-list evolution, the Slicing *Slice Static-list Forward AND,* on the FM Figure 2, will identify the following Figures 4 (a, b, c) which might be selected, modified, deleted, etc.

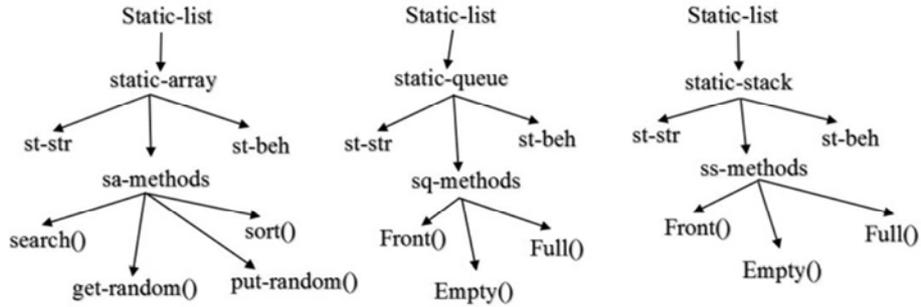

***Figure 4.** (a, b, c): The result of Slice Static-list Forward AND.*

As side effect to static-list evolution, the Slicing *Slice Static-list Forward OR static-queue,* on the feature model Figure 2, will identify the following Figures 5 which might be selected, modified, deleted, etc.

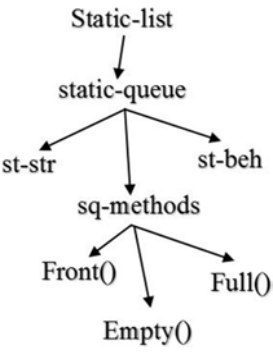

***Figure 5.** The result of Slice Static-list Forward OR static-queue.*

## 3. Results Discussion

The implementation environment of this work methodology requires a strongly typed object-oriented programming language and graph abstract data type. A strong combination between the FM, the reverse engineering, and feature slicing model should be guaranteed. Adding, an extension to an existing Object-Oriented Programming Language might be required to handle the added concepts of Feature-based Reverse Engineering.

Relying on the previous study on feature-based modelling and Reverse Engineering techniques, a comparison between the proposed methodology (including its supporting mechanisms) with similar recent works is presented below. This comparison is based on the following criteria: (1) Methodology supporting FM reverse engineering, (2) FM adaptation, (3) Supporting pattern recognition, (4) Slicing technique, and (5) Supporting tools. The studied similar works do not satisfy the criteria 1, 2, and 3. Some ones satisfy the criteria 4 but a mathematical way and some others satisfy the criteria 5. However, the proposed methodology covers all the first four criteria, but not the $5^{th}$. It satisfies the $4^{th}$ criteria in a graphical way, which better (than the mathematical one) for software requirements maintainers.

## 4. Conclusion

According to the previous works, the current approaches in FM Reverse Engineering encounter some insufficiencies, where the methodology is informal, not sufficiently adapted FM, no pattern for Feature Model and slicing technique are mathematically-based. This work consequently proposed a formal methodology, a uniform and adapted FM supporting reverse engineering concepts and algorithms, a feature pattern and pattern recognition algorithm and graph based slicing technique.

This work dealt only with feature-based requirements Reverse Engineering. Two feature patterns were introduced, the prospection of others feature patterns will be valuable. The slicing was limited to AND and OR relations, the study of others relations is important. What is about feature-based requirements reengineering modelling? and self-adaptation? The answers to these open questions will be challenging in this domain.

American Journal of Software Engineering and Applications 2019; 8(1): 1-7    7